\documentclass{article}
\makeatletter

\def\compoundrel#1\over#2{\mathpalette\compoundreL{{#1}\over{#2}}}
\def\compoundreL#1#2{\compoundREL#1#2}
\def\compoundREL#1#2\over#3{\mathrel
         {\vcenter{\hbox{$\m@th\buildrel{#1#2}\over{#1#3}$}}}}
\makeatother

\makeatletter
\leftmargin\leftmargini
\newdimen\Parindent\newdimen\Parskip
\Parindent=\parindent\Parskip=\parskip
\def\@oddhead{}\def\@evenhead{}
\def\@oddfoot{\rm\rightmark \hfil Page \thepage}
\def\@evenfoot{\rm\leftmark Page \thepage \hfil}

\newdimen\Parindent\newdimen\Parskip
{\Parindent=\parindent\Parskip=\parskip
\parindent0pt\parskip3mm\columnsep11mm}
{\parindent\Parindent\parskip\Parskip}
{\Parindent=\parindent\Parskip=\parskip
\parindent0pt\parskip3mm\columnsep11mm}
{\parindent\Parindent\parskip\Parskip}

\gdef\abstract#1{\gdef\@abstract{#1}}

\def\maketitle{\par
 \begingroup
 \setcounter{footnote}{0}\setcounter{page}{1}
 \let\save@thefootnote=\thefootnote
 \let\save@makefnmark=\@makefnmark
 \def\thefootnote{\fnsymbol{footnote}}
 \def\@makefnmark{\hbox
 to 2mm{$\m@th^{\@thefnmark}$\hss}}
\kern1mm 
\@maketitle
 \@thanks{}%
 \endgroup
 \let\maketitle\relax
 \let\@maketitle\relax
 \gdef\@thanks{}\gdef\@author{}\gdef\@title{}\let\thanks\relax}

\def\endtitlepage{
\@thanks{}%
\let\thefootnote=\save@thefootnote
\let\@makefnmark=\save@makefnmark
\setcounter{footnote}{0}\let\maketitle\relax\vskip 3mm\upshape
\gdef\@author{}\gdef\@title{}\global\let\thanks\relax
\global\let\@thanks\relax}

\def\@maketitle{\vbox to 40mm{\hsize\textwidth
 \linewidth\hsize \vfil \centering
 {\vskip 1em  \Large\bf \@title \par} \vskip 2em
{\begin{center} \large\bf \@author\end{center}\par}\vfil}
\hsize\textwidth \linewidth\hsize
\vskip 1em
\begin{center} ABSTRACT \end{center} \par
\begin{center}\parbox{140mm}{\@abstract}\end{center} \vskip 3em }

\renewcommand{\section}{\@startsection
{section}
{1}
{\z@}
{-1.0\baselineskip}
{0.1\baselineskip}
{\large\bf}}%
\renewcommand{\subsection}{\@startsection
{subsection}
{2}
{\z@}
{-1.0\baselineskip}
{0.1\baselineskip}
{\large\bf}}%
\renewcommand{\subsubsection}{\@startsection
{subsubsection}
{3}
{\z@}
{-1.0\baselineskip}
{0.1\baselineskip}
{\normalsize\bf}}%
\renewcommand{\paragraph}{\@startsection
{paragraph}
{4}
{\z@}
{-1.0\baselineskip}
{0.1\baselineskip}
{\normalsize\bf}}%
\renewcommand{\subparagraph}{\@startsection
{subparagraph}
{4}
{\z@}
{-1.0\baselineskip}
{0.1\baselineskip}
{\normalsize\bf}}%
\makeatother
\begin{document}
\title{Influence from the Future\thanks{Presented by H.\ B.\ Nielsen;
				the manuscript has
                                evolved appreciably since
                                the talk was given.}}
\author{Holger Bech Nielsen$^1$ and
	Colin Froggatt$^2$ \\
        {$^1$\it The Niels Bohr Institute,
	2100 K\o benhavn \O, Denmark}\\
	{$^2$\it Glasgow University,
        Glasgow G12 8QQ, Scotland}\\
\vspace{0.3cm}
{\rm To be published
in the Proceedings of the Fifth Hellenic
School and Workshops on Elementary Particle Physics, Corfu,
3 - 24 September 1995.}
       }
\abstract{We argue that some of the parameters in the laws of Nature would
be well understood, under the assumption that there
is an influence from the future as well as from the past, in the sense that
the principle of locality (or causality) is not valid at the fundamental
level. However, locality is supposed to be broken only in the mild way
that all the non-local influence comes from integrals over
the whole of space-time (including the past as well as the future)
and acts at all moments and all places with the same effect. Thus
the observable effects of the lack of locality can only be seen in the
constants of Nature. Our cleanest prediction is the Higgs boson mass being
$M_H=149 \pm 26$ GeV, which in addition assumes that the pure Standard
Model is valid (i.~e.\ no new physics) until the Planck scale. Adding
the assumption that the two
vacuum states needed in our model come about naturally requires
a strong first order transition between them;
together with the assumption
of the Planck units being fundamental, this leads
us to also predict the mass of the top quark, $M_t = 173 \pm 5$ GeV,
and a more precise value of the Higgs boson mass, $M_H=135 \pm 9$ GeV.
With the assumption of our favourite gauge group beyond the Standard Model:
the anti-grand unified group SMG$^3$
(three copies of the Standard Model gauge group,
one for each generation, first appearing close to the Planck scale),
it is also possible to obtain
impressively good predictions for the three fine structure constants in the
Standard Model. We also discuss the
extension of the SMG$^3$ group by an extra abelian factor
$U(1)_f$ and the fermion mass matrices in this model.
Other vacua than the present one might appear in the future due to
human activity. It is even
suggested that life itself could be the miracle needed
to avoid a time machine paradox like the matricide paradox:
life and we ourselves were made in order
to produce a vacuum bomb inaugaurating the next vacuum.
The breaking of locality is really of the same mild character as suggested by
baby universe theory, so the present work
may be considered as a development of
baby universe theory.}

\maketitle
\section{Introduction}\label{sec:intro}
In the formalism of general relativity there is no obvious obstruction to the
appearance of wormholes and baby universes \cite{baby}, or other
topological configurations that
easily cause communication of a type incompatible with the
validity of a causality or locality principle.
It may be technically difficult and even difficult
for fundamental reasons---namely the need
for negative energy density---to
produce large scale wormholes. However at the Planck scale
$\sim 10^{-33}$ cm there can be huge quantum fluctuations
and even the topology of space-time is expected to fluctuate,
so it is hard to see
how the locality threatening space-time structures could be prevented.
Given quantum mechanics and gravity, it seems almost impossible to construct
a theory without space-time foam and non-locality.
In baby universe theory it is speculated that the typical
such locality violating configuration, namely a handle
connected to a weakly curved (relative to the
handle size) four dimensional space-time,
is to be integrated over all possible ``points'' of attachment
for the two ends of the handle. From this integration, it follows that
the effect of the handles (i.~e.\ of the baby universe exchanges) are the
same all over space-time. It is understandable that the effect is, in this
way, smeared out to be the same all over space-time, from the fact that a
baby universe leaving a smooth universe is prevented from taking
energy or momentum with it. In Einstein's theory of general relativity,
the information about energy and momentum is to a large extent
stored ``safely'' in the gravitational field far away
from where the energy or momentum is really
situated. Heisenberg's uncertainty relation may then be
used to argue that, since
the leaving baby universe does not take away any 4-momentum,
the probability amplitude for its place of
departure, i.~e.~the ``point'' of attachment of the
handle, is the same all over space-time.
In this way the non-local effect becomes
of the mild type that one can never switch off and
must be the same all over space-time.
Thus, after integrating out the baby universes,
the only residual effect of the non-locality
is to change the coupling constants in the low energy effective
Lagrangian density.
In this way it is not so easy to exclude, on empirical grounds,
the existence of such non-local effects.

In the present work it is actually our point to
argue that the observed values of the coupling constants
support the belief that non-local effects do
exist and that, indeed, the future does influence us
via such baby universe
exchanges (or perhaps simply due to a fundamental lack of locality).
If, in the spirit of the project of Random Dynamics \cite{book},
one would like to ``derive'' all the
known laws of Nature in some limit from an essentially random fundamental
dynamics,
then one should indeed at
some level investigate the possibility that the principle of locality
is not valid fundamentally, but rather to be hopefully derived in some
limit. Anyway the main subject of the present article is to use
what we believe is a likely consequence of the just suggested
baby universe-like mild form of locality-breaking, namely the multiple point
criticality principle \cite{glasgowbrioni},
roughly meaning that the vacuum has many essentially degenerate
phases. It is the fine-tuning of the various parameters of the Standard
Model, needed to achieve such degeneracies of the different vacuum states,
that makes up the messages from the future, as well as the past, and which we
want to claim supports the picture sketched above of mild locality breaking.
We shall see below that, if our prediction of the
Standard Model Higgs boson
mass of 149 $\pm$ 26 GeV turns out to be correct experimentally,
it would be an especially clean test of our claim
and it is of course a true {\em pre}diction.

In section 2 we discuss the Standard Model effective Higgs
potential and the two main conditions we impose on it,
in order to obtain our predictions for both the top quark
and Higgs boson masses. The motivation for these two
conditions is provided by our assumption that there is a breakdown
of locality at the Planck scale. Non-locality and the resolution
of the time machine paradox, by requiring the presence of
two different vacua in different space-time regions, are
discussed in section 3.

In section 4 we argue that it is only likely that the time machine paradox
works provided the non-local effect, sensitive to the vacuum being one
or the other, is quite strong. This suggests that
the vacuum
expectation values for the Higgs field, in the two
degenerate phases, deviate
from one another by
an amount of the order of unity in Planck scale units.
Our calculation of the
predictions of the top quark mass and Higgs boson mass,
from this requirement, is then presented in section 5.
In section 6 a slightly different, but presumably basically equivalent,
formulation of the model is presented, using
an ice-water microcanonical ensemble as
an analogue, thereby showing that our ideas are
not without a kind of precedence in Physics.

In section 7 we briefly describe other work using the same
multiple point criticality principle, together with the assumption of
the anti-grand unified gauge
group $SMG^3$, or perhaps the closely related
group $SMG^3\times U(1)_f$, leading to fine-tuned values of
the fine structure constants. In fact
the three gauge couplings (fine structure constants) in the
Standard Model, $\alpha_1$ , $\alpha_2$, and $\alpha_3$, are
successfully predicted to
around ten percent accuracy at the Planck scale.
In section 8 we discuss how the extra chiral gauge quantum numbers
of the $SMG^3\times U(1)_f$ model can be used to obtain a
realistic fermion mass and mixing hierarchy.
Finally, in section 9, we give our conclusion, summary and consolation.

\section{Effective Higgs Potential $ V_{eff}(\phi)$}

It is well-known that in a renormalisable quantum field
theory the bare (classical) potential should only have terms
involving field factors up to mass dimension
four (in four space-time dimensions)
and thus, in the Standard Model, must be of the form:
\begin{equation}
V(\phi) \; = \; \frac{1}{2}m_{0H}^2 |\phi |^2 \;
+ \; \frac{1}{8}\lambda_0 |\phi |^4
\label{bare}
\end{equation}
Since the Higgs field $\phi$ has $ U(1)$ weak hypercharge
as well as being a weak isospin doublet, it must occur in the
combination $|\phi|^2$. However, when quantum fluctuations of
fields are taken into account and one asks for the energy density in
a state with a prescribed average value of the Higgs field
(in a prescribed gauge), one cannot simply use the bare potential
formula eq.~(\ref{bare}), but rather the effective potential \cite{sher}
given to the first approximation (with one-loop corrections) by:
\begin{equation}
V_{eff}(\phi) = \frac{1}{2}m_{0H}^2 |\phi |^2
+  \frac{1}{8}\lambda_0 |\phi |^4
- \frac{1}{2}\mbox{Tr}\,\log\,[\delta^2 S/\delta \phi \delta \phi\,
(\delta^2 S/\delta \phi \delta \phi|_{\phi=0})^{-1}]
- \mbox{Tr}\,\log\,[\mbox{for other fields}]
\label{simplest}
\end{equation}
where $S$ is the classical action.
The effective potential $V_{eff}(\phi)$ may really
be thought of as taking into account
the energy density due to zero point fluctuations
in the Higgs field $\phi$
around its central (= average) value, as well as how the
zero point fluctuation energy density of the other fields vary
as a function
of $\phi$; this is at least so for
static field configurations. If, as we shall actually assume,
the effective potential, calculated according
to this formula eq.~(\ref{simplest}), has
two minima, at $\phi_{\rm{min}\; 1}$
and $\phi_{\rm{min}\; 2}$,  with
a hill in between, a prescribed average value for the field $\phi$ in the
interval of such a hill would (to get minimum energy density) be
realized as
an inhomogeneous mixed state. The Higgs field $\phi$
in this mixed state would take on different values
$\phi(x) \simeq \phi_{\rm{min}\; 1}$
and $\phi(x) \simeq \phi_{\rm{min}\; 2}$
in various regions of space-time,
rather than by having the
field closely fluctuating around a single $\phi$-value in the hill
interval. This means that if
the effective potential is defined
precisely
as the the minimal energy density obtainable
in a state, subject to
the constraint of having a prescibed average value for
$\phi$, then the expression in eq.~(\ref{simplest}) is not correct
for  $\phi_{\rm{min}\; 1} < |\phi| < \phi_{\rm{min}\; 2}$.
With this precise definition, the effective potential
will always be a convex function.
However, under conditions where the tunnelling from
one such mentioned minimum at $\phi_{\rm{min}\; 1}$
to the other minimum at $\phi_{\rm{min}\; 2}$ is
strongly suppressed,
the effective potential, as defined in eq.~(\ref{simplest}),
plays an important role for times small compared to
the large tunnelling time. Therefore, in the
present article, we shall use
the not necessarily convex expression
eq.~(\ref{simplest})---or rather the
more accurately calculated expression
discussed below--- for $V_{eff}$, ignoring the
possibility of averaging over fields
which take on wildly different values in different regions.
In the following sections, we shall
discuss this ``averaging'' in a more
``physical'' way, talking about vacuum bombs, the future and
destiny of humanity etc., rather than just hiding
it in a formal definition of the effective potential.

One can, of course, improve the lowest order (one-loop)
effective potential eq.~(\ref{simplest}), by using
higher order corrections.
An especially efficient method of
improving $V_{eff}$ is to make use
of the running coupling constants---by far the most
relevant for our work is the Higgs field self-interaction coupling constant
$\lambda(\mu)$---as calculated  by integrating up the renormalisation
group equations for the various couplings in the Standard Model.
A large part of the trace-log corrections in the one-loop approximation
formula eq.~(\ref{simplest}) can actually be included, by just using the
form of the bare potential eq.~(\ref{bare}),
but inserting the running coupling constants as coefficients,
identifying the scale parameter $\mu$
with the field value $|\phi |$. Taking into account the difference
between the bare and the running (= renormalised) couplings
in  eq.~(\ref{simplest}), one can
obtain the renormalisation group improved effective potential. But
just using the running coupling(s) in the bare form of the potential
is already quite
good, and essentially what we shall use in our own discussion. Only for
the purposes of
high accuracy calculations is the true renormalisation group
improved one loop effective potential really needed.

That is to say a rather good approximation is already achieved
by just taking:
\begin{equation}
V_{eff}(\phi) \; = \;
\frac{1}{2}m_{H}^2(\mu = |\phi |)\, |\phi |^2 \; + \;
\frac{1}{8}\lambda (\mu = |\phi | )\, |\phi |^4
\end{equation}
The concept of a running Higgs mass squared $ m_H^2(\mu )$ may only
be reasonably defined by excluding
the quadratic divergences
as causing any scale change.
However,
the term with the Higgs mass squared coefficient
becomes unimportant for Higgs fields large
compared to the weak scale, and that is where we have our main interest.
The running $\lambda ( \mu) $ is easily computed by means of the
(first order) renormalisation group equations:
\begin{equation}
16\pi^2\frac{d\lambda}{d\ln\mu} =12\lambda^2 +
3\left(4g_t^2 - 3g_2^2 - g_1^2\right)\lambda +
\frac{9}{4}g_2^4 + \frac{3}{2}g_2^2g_1^2 + \frac{3}{4}g_1^4 - 12g_t^4
\label{rgelam}
\end{equation}
Here the $g_i(\mu)$ are the three Standard Model
running gauge coupling constants,
discussed further in section~\ref{sec:fine},
and $g_t(\mu)$ is the top quark running Yukawa coupling constant,
which satisfies the renormalisation group equation:
\begin{equation}
16\pi^2\frac{dg_t}{d\ln\mu} = g_t\left(\frac{9}{2}g_t^2 - 8g_3^2
- \frac{9}{4}g_2^2 - \frac{17}{12}g_1^2\right)
\label{renormalizationgr}
\end{equation}
Because the top quark
Yukawa coupling is of order unity, while
the other Yukawa couplings are very small,
it is only the top quark Yukawa coupling
that is significant for the renormalisation group development
of the Higgs self-coupling. So we ignored the other quark and lepton
Yukawa couplings, including transition Yukawa couplings (quark
mixing angles).
For large values of $|\phi |$ the effective potential will
approximately be given as $V_{eff}(\phi) \approx \frac{\lambda
(|\phi |)}{8} |\phi |^4$, where $\lambda(|\phi |)$ is determined
by solving the differential equations (\ref{rgelam}) and
(\ref{renormalizationgr}).
In practice we used the second order renormalisation
group equations.

With the bare potential form eq.~(\ref{bare}), it is only
possible to have two minima as a function of $|\phi | $
by having one of them at $|\phi| =0$.
However, for sufficiently large $g_t$,
it is possible to arrange  the renormalisation group improved
effective potential to have
two minima, both of which correspond to strictly positive
values of $|\phi |$. With the well-known
vacuum expectation value of the Higgs field, $<\phi> $ = 246 GeV,
the top quark Yukawa coupling constant must have a value
corresponding to a top quark mass of about 90 GeV or above,
in order
to obtain two minima with
strictly positive $|\phi|$ (norms).
If there are indeed two minima,
for a top quark mass larger than 90 GeV, the realistic picture
is that the minimum corresponding to the experimentally observed
properties of weak interaction physics is the smaller $|\phi|$-valued
one, $\phi_{\rm{min}\; 1}$.
The Higgs boson mass that will be observed
is, of course, what we
could call a renormalised Higgs mass and is very closely, but
not completely exactly, given by the second derivative of the effective
potential
at $\phi_{\rm{min}\; 1}$  = 246 GeV.
It is the
second derivative in the radial direction that is to be used here.
The point of the
further small correction to obtain the true Higgs pole mass is that
the second derivative of $V_{eff}$ really gives what
one might call the Higgs mass squared
when the four momentum of the Higgs boson is
zero, but the on mass-shell condition does not of course
correspond to precisely zero four momentum.
This correction is quite calculable and not big.

It is reasonable to think of the free parameters, corresponding to the
relevant bare couplings $g_{0t}$, $m_{0H}^2$, $\lambda_0$ and the
three Standard Model fine structure
constants, as to be given by fixing:

A) the renormalised fine structure constants $\alpha_i=g_i^2/(4\pi)$
and the value of the Higgs field norm $| \phi | $
for the phase in which we live,  $\phi_{\rm{min}\; 1}$ = 246 GeV;
these parameters are already very well-known experimentally---the latter
vacuum expectation value being essentially
an expression for the Fermi-constant of weak interactions;

B) the two less well-known parameters represented by the Higgs boson and
top quark masses (with $<\phi>$ given, the Higgs boson mass is essentially
related to $\lambda$ while the top quark mass is related to $g_t$).

The main results presented here are
based on the following numerical observation.
If the parameters listed under A are fixed by experiment
and the two remaining parameters---the Higgs boson and top quark
masses---are adjusted so that the effective
Higgs potential gets two equally deep minima, the second of which
has a Higgs field vacuum expectation value of the order
of the Planck scale, then the required masses are quite
realistic from an experimental point of view. The
needed top quark mass turns
out to be $173$ GeV, accidentally just the value first estimated
by the CDF experiment at FNAL. The corresponding Higgs boson mass is 135 GeV
and is at least consistent with all present bounds, as well as with indirect
estimates from LEP measurements and radiative corrections.
Figure \ref{fig:veff} illustrates the two features
which, we thus claim, are quite
compatible with
experiment and the pure Standard Model being valid up to
the Planck scale $M_{\rm{Planck}}$.

\begin{figure}[t]
\vspace{11.2cm}
\caption{Symbolic illustration of the two main requirements on the
effective potential $V_{eff}(\phi)$ which we observe leads to
acceptable numerical values for the top quark and Higgs boson masses:
1) Two equally deep minima, 2) achieved for $|\phi|$
values differing, order of magnitudewise, by
unity in Planck units.
}
\label{fig:veff}
\end{figure}

We now wish to show that our idea of the existence of two
Standard Model vacua, having the same energy density and with
the difference in their Higgs field vacuum expectation values
being of order unity in fundamental (Planck) units, can be
motivated by the consideration of non-local effects.
So, in the next two sections, we take up our assumption that
the otherwise so well-established principle of locality
is actually fundamentally violated!

\section{Vacuum Bomb and Time Machine}
The major assumption in our work---in addition to believing in the pure
Standard Model essentially all the way to the Planck scale---is that
the principle of locality (or causality ) is broken at the fundamental level!
This may at first seem a very strong, and even obviously wrong, assumption.
However, except at very short distances of the order of the Planck scale,
we assume that it is {\em only broken in the mild way that there is an
effect which is the same at all points in the space-time manifold, and that
this effect depends on an averaging over all space-time}. This translational
invariant non-local effect is much less easy to reveal empirically, if it
existed, than an effect that could be switched on and off. It will namely
be conceived of as a modification of the parameters in the laws of Nature,
the ``coupling constants'' we may call them, and our non-locality in this
mild form only means that the coupling constants depend on what has
happened in the past and what will happen in the future (in the sense of
an average over all of space-time).
Since it is hard to know ``what the coupling constants
should have been '' without such non-locality effects, it is hard to see
that there was any modification of them due to effects from
the past and the
future. It is, however, not a priori impossible, since it could happen
that the non-local effects leave such a characteristic signal in the
pattern of coupling constants,
that we would become (psychologically and scientifically) convinced
of the existence of such a non-local mechanism. In fact that
is what we want to say: With only very few extra assumptions, the effects
of non-locality predict the Higgs particle mass to be 135 GeV, the
top quark mass to be 173 GeV, the running Standard Model
fine structure constants at the Planck scale correct within 10\%,
zero cosmological constant
etc.~! We in fact, to a large extent, see such a pattern
and so perhaps the reader really ought to believe that
locality is broken in the mild way we suggest.

It should be remarked that this mild way of breaking locality---and even
the idea of breaking locality---is taken over from  baby universe theory.
Baby universe theory \cite{baby} is the
study of the effect of small handles---exchange of baby universes---on
an otherwise smooth space-time manifold. Since the
attachment points on the smoother
space-time manifold are integrated over, each handle provides effects which
are the same all over the smoother manifold.
That is to say the baby universes
precisely provide breaking of locality only in the mild way we just described,
giving the same effect all over space-time.
You may thus consider our non-locality a (slight)
generalization of baby universe theory. We consider it a generalization,
because we do not exclude the possibility that the fundamental laws of Nature
lack the principle of locality fundamentally and not just due to
some baby universes popping up, in an otherwise a priori local looking
theory of quantum gravity. Locality plays a somewhat roundabout
r\^{o}le in baby universe theory:
In quantum gravity one would work with a Lagrangian
that is required to be local, in the sense of being
an integral over four dimensional space-time
of a Lagrangian density only depending
on finite order derivatives of fields at the point in question
(similarly string theory is described by a local action).
Then baby universes appear and locality
gets really broken, but now---hopefully---it turns out to be so
mildly broken that, after all, it can only be seen
in the coupling constants, as we described above, and locality is
effectively restored.

This picture is a bit strange philosophically: The scientist has put the
a priori locality assumption  of the string theory, or of the quantum gravity
action written as a space-time integral, into the models
presumably based on the
empirical observation of locality; but the empirical observation of locality
is strongly based on the fact that the violation of locality (by
baby universes) is only {\em mild}, i.~e.\ the same everywhere so that it only
appears in the coupling constants. The basic input of locality at the
very fundamental level is not sufficient! So we might as well say
that assuming locality broken quite a priori at the
fundamental level is preferable.
In the spirit of Random Dynamics one would take locality to be broken
fundamentally and indeed we have argued \cite{book} that, assuming general
reparameterisation invariance, it can easily turn out only to be broken in
the mild way; thus we conclude that we really should never
have assumed locality at the fundamental level.
Anyway it does not matter in the present work,
we just need the breaking in the mild way for some reason or the other.

Now in the previous section we saw that, with an appropriate
effective Higgs potential, there is the possibility for several
``vacuum states'' to exist. The presence of matter,
of the type and density which we observe in our Universe today,
should be considered as only tiny modifications of the
vacuum fields, compared to the magnitude
of the field modifications needed to shift from one vacuum to
another one; the important events, from the physical point of
view, are not whether some country gets a new president or so
but rather whether we get a new vacuum. It is
the presence of different vacua in different space-time regions
which is by far the most important feature in space-time, as it is
responsible for the effects of non-locality resulting in
contributions to the ``coupling constants'' (the parameters
in the laws of Nature). That is to say:
the messages from the future, which we can realistically hope to
get in our picture, relate to information about the vacuum in
coming epochs. We shall refer to this influence
from the vacuum in the future as a type of time machine.

Let us, for simplicity, imagine that we have just the possibility of
the two vacua achievable by putting the Standard Model
Higgs field expectation value equal to either
$\phi_{\rm{min}\; 1}$ or  $\phi_{\rm{min}\; 2}$,
as in figure \ref{fig:veff}. In reality we may believe
that, by including other degrees of freedom beyond those of
the pure Standard Model or by considering non-perturbative
effects on a lattice, we could find many other possible
vacua; however that does not matter so much for
understanding the principal idea. Let us further suppose
that one specific phase---our vacuum, the one we live in
now with $<\phi> = \phi_{\rm{min}\; 1}$---came
victoriously out of the Big Bang. That is to say it
came to be the realized phase all over three-space.
If the energy density in the various vacuum-candidates
were very different it is almost unavoidably the case that
the lowest energy phase would win as the cooling of the
Universe proceeded, since the higher energy density phases
could relatively easily decay into the lowest energy
density phase. However, if the energy differences, for some
reason or the other, happened to be small,
then it could be that a ``vacuum'' other than the one of
lowest energy density was the most stable at
the high temperature prevailing in the time shortly
after the Big Bang. Such a ``vacuum''
could then have pushed the other ones away and, by being metastable,
survive for a long time---perhaps even an infinitely long time.
This possibility opens up
the scenario of a vacuum bomb being constructed and causing
the transition of space from the metastable phase into
the stable one.  The moment when that happens, of course,
in principle depends on details of the development of the Universe
and on politics etc.\ if human beings are involved in triggering the
transition off.  Let us denote this moment of vacuum transition
by $t_{\rm{ignition}}$; then the various coupling constants,
such as e.~g.~the bare Higgs mass, will depend in a rather
smooth way on this ignition-time $t_{\rm{ignition}}$
due to non-local effects (in the mild form).
But now there is the possibility of a paradox, much like
the matricide paradox (this is a paradox that can be reached
if one possesses a time machine and uses it for going backward in
time to kill one's own mother while she is still a child;
at first there seems to be no possible solution as to what
can happen, because all possibilities seem contradictory!).
It could very likely happen---with say 50\%
probability---that the non-local effect is such that a late
ignition of the truly stable vacuum leads to a value for
the Higgs mass, which would give the stable vacuum a much
lower energy density than the metastable one; this in turn
would imply that the ignition ought to have
taken place very shortly after the temperature of the Universe
had fallen sufficiently that the truly stable vacuum was in fact
also the one with the lowest free energy. While, on the other hand,
an early transition would lead to the high temperature favoured
vacuum having such a low energy density that it would actually be
the truly stable one (while the other one would have higher
energy density and would never get competitive).
In this case we have indeed obtained a paradox:
If the ignition time $t_{\rm{ignition}}$ is early in the
evolution of the Universe the Higgs mass will be such as to make it
late or totally impossible, and if it is late the non-local
effects will make it early; a contradiction in each case!

\begin{figure}[t]
\vspace{10.5cm}
\caption{The ignition time $t_{\rm{ignition}}$ for a vaccum bomb and
the development
of the effective Higgs potential $V_{eff}(\phi)$ as a function of
the bare Higgs mass parameter $m_{0H}^2$. The solid line labelled
``normal physics'' shows the usual dependence of $t_{\rm{ignition}}$
on $m_{0H}^2$. Whereas the other four solid lines illustrate
possible strong effects of non-locality. The intersection of a
``non-locality'' curve with the ``normal physics'' curve
occurs for a value of  $m_{0H}^2$ for which the minima of
$V_{eff}(\phi)$ are approximately degenerate.
}
\label{fig:ignition}
\end{figure}

Acording to Novikov and others \cite{novikov,thorne},
in these paradoxical cases
with time machines, one should find one or more self-consistent solutions.
In the case of billiard balls passing through a wormhole and
traversing a closed timelike curve, say, one can indeed find such
solutions, although they typically have the character
of being {\em fine-tuned}, such as requiring that one ball just
barely touches another (or itself returned from the future);
the bullet shot at one's mother could give a superficial wound
only. We can also express this by saying that the existence
of time machines would imply miracles!

In figure \ref{fig:ignition} we have shown the expectation from
normal physics
that the time of ignition largely speaking increases as the
bare Higgs mass squared $m_{0H}^2$ increases, because with
$m_{0H}^2$ increasing the energy density of the
other large $\phi = \phi_{\rm{min}\;2}$
``vacuum'' increases and
therefore becomes more difficult to reach from our vacuum.
The small graphs under the main abscissa axis show the
corresponding development
of the effective Higgs potential $V_{eff}(\phi)$ as a function of
$m_{0H}^2$, with the two minima
of  $V_{eff}(\phi)$ denoted
by ``us'' and ``other'' respectively.
Eventually $m_{0H}^2$ becomes so
large that it is the high $\phi_{\rm{min}\;2}$ vacuum which is unstable
and we shall then of course never find any transition, assuming
that our low $\phi_{\rm{min}\;1}$ vacuum won the competition
of coming out of the Big Bang.
Thus the ignition time $t_{\rm{ignition}}$ diverges when
the value of $m_{0H}^2$
is such that the two vacua become degenerate.
This is so to speak the normal
dependence between  the ignition-time $t_{\rm{ignition}}$
and the bare Higgs mass squared $m_{0H}^2$.
However, with non-locality there is also a (more mysterious)
non-local effect and that could easily, as we suggested,
go in the opposite direction, i.~e.~make $m_{0H}^2$ a
decreasing function of $t_{\rm{ignition}}$.
Four possible non-locality curves are drawn in
figure \ref{fig:ignition}, illustrating what the
non-local effect could be.
The only self-consistent
solution is where the non-locality and normal physics
curves intersect. In each of the four cases,
the self-consistent solution has
a value for the bare Higgs mass squared $m_{0H}^2$
such that the two vacua are nearly degenerate.
That is to say that a ``time machine miracle'', of the same
sort that makes a billard ball running into a time machine
(wormhole) and knocking its ``younger'' self
away from the arranged orbit into the wormhole
actually just graze itself so that a consistent
development occurs, will also make the time machine for mild
lack of locality  become consistent. Here the miracle is that
the vacua become just degenerate: our so-called multiple point
criticality principle. This means that the vacuum in
which we live---``our'' vacuum---lies on the vacuum stability curve
\cite{sher,drtjones} and hence the Higgs mass is determined as a
function of the top quark mass. In particular for a
top quark mass \cite{CDF} of  $180 \pm 12$ GeV, it follows from
recent calculations of the vacuum stability curve
\cite{shervs,isidori,casas} that the Higgs pole mass is
predicted to be $149 \pm 26$ GeV.

\section{Likelihood of Mechanism, Strong First Order Transition Requirement}

At first one might think there is no restriction on what
values, $\phi_{\rm{min}\; 1}$ and  $\phi_{\rm{min}\; 2}$,
the Higgs field could take in the two minima, but
really there is an argument that the difference must
be rather ``large ''. The point is that if the difference,
$\phi_{\rm{min}\; 2} - \phi_{\rm{min}\; 1}$,
in Higgs field
values for the two vacua
is small, the effect of say the future acting back
on the present, depending on which vacuum is present in the future, will
be small too. If that is the case there will very likely not be any
time machine paradox in the first place and, thus,
also no miraculous solution
will be needed. Then the Higgs mass parameter $m_{0H}^2$
only depends very weakly on the
time of ignition of the new vacuum and
it is, to first approximation, as if there was no time machine at all.

What now means ``large'' or ``small'', for such an influence of
say the future on the past and present etc.? We shall only estimate
it in a dimensional way.
Since we think that the basis of
our model is quantum gravity---baby universes and the
like---it is natural to assume that the fundamental units are to be
taken as the Planck units, i.e. the units based on Newton's
gravitational constant $G$, the speed of light $c$ and the
Planck constant $\hbar$: $M_{\rm{Planck}}\simeq10^{19}$ GeV.
Then we can from dimensional arguments expect that a ``large''
effect will appear when the field difference,
$\phi_{\rm{min}\; 2}- \phi_{\rm{min}\; 1}$,
is large compared to
one in the Planck units for that field;
this means that the ``phase transition'' between the
two vacua is strongly first order.
Since now ``our'' vacuum has an
extremely small Higgs field expectation value,
$\phi_{\rm{min}\; 1}$, the difference is roughly the
same as the other vacuum field value,  $\phi_{\rm{min}\; 2}$.
Thus we must have $\phi_{\rm{min}\; 2}$ at least of order
$M_{\rm{Planck}}$ to make the whole scheme work.
On the other hand it becomes natural that  $\phi_{\rm{min}\; 2}$
is not very much larger than  $M_{\rm{Planck}}$,
when we assume---as we do---that the Planck units are
the fundamental ones. Thus we end up suggesting that indeed the
second minimum must be at the Planck scale:
$\phi_{\rm{min}\; 2} \simeq M_{\rm{Planck}}$. It is this assumption that
then leads to the top quark mass prediction of 173 GeV,
as we shall see in the next section.

Whatever the arguments for the second minimum to be at the Planck scale,
it can only be true in order of magnitude, because the very concept
of the Planck scale is only defined via dimensional arguments, making
sense modulo some factor of the order of unity. Our prediction of the
top quark mass can therefore be no more accurate than
a calculation with an order of magnitude uncertainty in  $M_{\rm{Planck}}$.
It is remarkably good luck for our calculation
that the dependence on the value used for $M_{\rm{Planck}}$
turns out to be very weak indeed.

For dimensional reasons, the predicted top quark Yukawa coupling must depend
upon the ratio of the weak scale to the Planck scale.
Now it is well-known \cite{corfu,pendleton,hill} that
the Standard Model renormalisation group equations (\ref{rgelam}) and
(\ref{renormalizationgr}) have an
infrared quasi-fixed point, when the top quark Yukawa coupling
and the Higgs self-coupling are considered as running, but the
fine structure constants as being essentially
scale independent. As the renormalization point $\mu$
goes down towards the infrared, $g_t$ and $\lambda$ approach this
approximate infrared fixed point exponentially as a function of
$\log \mu$. But that means that the approach goes as some power law as
a function of $\mu$ for reasonably small $\mu$ (such that we
get dominance by
the infrared point, but not so low $\mu$ that the approximation in which
there is a fixed point is spoiled). The value of the running $g_t$ relevant
for our top quark mass prediction
is in a regime where this approximate fixed point
is a reasonable approximation and is approached from below.
Thus the deviation of the top quark mass from the fixed point
value obtained in the infra-red limit goes as
a power law function of the $\mu$ to Planck scale ratio. But then
it also goes as a power law as a function of the Planck scale,
when the weak scale is considered fixed. We looked at the numerical
curves and extracted a series of top quark masses (for given weak scale)
as a function of $\phi_{\rm{min}\; 2}$. It turned out
numerically, from these correlated top quark mass and
$\phi_{\rm{min}\; 2}$ values,
that the predicted top quark mass deviates from the quasi-fixed point value,
of ca. 230 GeV,
by a term going inversely as the 42nd root of the Planck scale
(relative to the weak scale).

To some extent this connection with the infrared quasi-fixed
point top quark mass
takes away some of the great impressiveness of our prediction, because
it means that in the very first approximation we just obtain the
fixed point value; this of course will be the result
in essentially any model having a large desert, so that the weak scale
is in the infrared, and an unsuppressed top quark Yukawa
coupling. Nonetheless
our prediction actually deviates from the fixed point value by about 25\%,
and agrees with experiment much better than if
we had just predicted a value
in the neighbourhood of the
infrared stable point. Indeed we see that the $42$nd root
converts an uncertainty of a factor of 10 in the Planck scale
to only an uncertainty of
$\frac{\ln 10}{42} \times 100\% \simeq 5\%$
in the  deviation between the fixed point value
and the predicted top quark mass.  This corresponds then to a couple
of percent uncertainty in the total top quark mass,
or of the order of $3$ GeV.

\section{Calculation of the Higgs and Top Masses}

As discussed in section 3,
the influence from the future in a non-local theory can easily
lead to the requirement that the vacuum should have degenerate
phases. This gives the condition that the Standard Model
renormalisation group improved
effective Higgs potential should take the same value
in two minima:
\begin{equation}
V_{eff}(\phi_{\rm{min}\; 1}) \quad =
\quad V_{eff}(\phi_{\rm{min} \; 2})
\label{eqdeg}
\end{equation}
One of the minima corresponds to our vacuum with
$\phi_{\rm{min}\; 1} = 246$ GeV and eq.~(\ref{eqdeg})
defines the vacuum stability curve. We are interested in the
situation when $\phi_{\rm{min} \; 2}\gg\phi_{\rm{min}\; 1}$. In
this case the energy density in our vacuum 1 is exceedingly small
compared to $\phi^4_{\rm{min} \; 2}$. Also, in order that
$\phi_{\rm{min}\; 1} = 246$ GeV, the coefficient
of $\phi^2$ in the effective Higgs potential has to be of order the
electroweak scale. Thus, in the other
vacuum 2, the $\phi^4$ term will a priori
strongly dominate the $\phi^2$ term. So we basically get the
degeneracy condition eq.(\ref{eqdeg}) to mean that, at the
vacuum 2 minimum, the effective coefficient $\lambda(\phi_{\rm{min} \; 2})$
must be zero with high accuracy.  At the same $\phi$-value the derivative
of the effective potential $V_{eff}(\phi)$ should be zero, because
it has a minimum there. In the approximation $ V_{eff}(\phi) \approx
\frac{1}{8}\lambda(\phi) \phi^4 $ the derivative of $V_{eff}(\phi)$
with respect to $\phi$ becomes
\begin{equation}
\frac{dV_{eff}(\phi)}{d\phi}|_{\phi_{\rm{min} \; 2}}
= \frac{1}{2}\lambda(\phi)\phi^3
+\frac{1}{8}\frac{d\lambda(\phi)}{d\phi}\phi^4
=\frac{1}{8}\beta_{\lambda} \phi^3
\end{equation}
and thus at the second minimum the beta-function (given
to first order by the right hand side of eq.~(\ref{rgelam}))
\begin{equation}
\beta_{\lambda} =
\beta_{\lambda}(\lambda(\phi), g_t(\phi), g_3(\phi), g_2(\phi),g_1(\phi ))
\end{equation}
vanishes, as well as $\lambda(\phi)$.
Here we used the approximation of the renormalisation group improved
effective potential \cite{sher},
meaning that we used the form of the polynomial classical
potential {\em but}\ with running coefficients taken at the renormalisation
point identified with the field strength $\phi$. We also do not distinguish
between the field $\phi$ renormalised, say, at the electroweak
scale and the renormalised running field $\phi(t) = \phi \xi(t)$ at another
scale $\mu(t)=M_Z \exp(t)$ where $ \xi(t) = \exp(-\int_0^t dt' \frac{\gamma}
{1-\gamma}) $. The reason is that, due to the Planck scale being only  used in
order of magnitude, we shall get uncertainties of the same order as
this correction. In fact the anomalous dimension $\gamma$ is of the order
of 1/100, making the difference at most of the order of our uncertainty.

The degenerate minima condition eq.~(\ref{eqdeg}) and the
associated vacuum stability curve have been studied for the
Standard Model in three recent
publications \cite{shervs,isidori,casas}.
Their results are slightly different but, within errors,
are each consistent with the linear fit
\begin{equation}
M_H = 135 + 2 (M_t -173) - 4 \frac{\alpha_3 - 0.117}{0.006}
\label{eqvacstab}
\end{equation}
to the vacuum stability curve, in GeV units.
When this degenerate minima condition eq.~(\ref{eqvacstab})
is combined with the experimental value
\cite{CDF} of the top quark pole mass, $M_t = 180 \pm 12$ GeV,
we obtain a rather clean prediction \cite{smtop} for the Higgs
pole mass:
\begin{equation}
M_H \quad = \quad 149 \pm 26 \; \mbox{GeV}
\label{eqmhiggs}
\end{equation}

If we now also impose the strong first order transition requirement,
discussed in the previous section, which takes the form:
\begin{equation}
\phi_{\rm{min}\; 2} \quad = \quad
{\cal O}(\rm{M}_{\rm{Planck}})
\label{eqstrong}
\end{equation}
we no longer need the experimental top quark mass as an input,
but rather obtain a prediction for both $M_H$ and $M_t$.
So we impose the conditions $\beta_{\lambda}=\lambda=0$
near the Planck scale,
$\phi_{\rm{min} \; 2} \simeq M_{\rm{Planck}}$,
and use the second order renormalisation
group equations to evaluate $g_t(\mu)$ and $\lambda(\mu)$
at the electroweak scale $\mu = \phi_{\rm{min} \; 1}$ \cite{smtop}.
A change in the scale
of the minimum $\phi_{\rm{min} \; 2}$ by an order of magnitude, from
$10^{19}$ GeV to  $ 10^{18}$ or $ 10^{20}$ GeV, gives a shift
in the top quark mass of
ca. 2.5 GeV. Since the concept of Planck units only makes
physical sense w.r.t.\ order of magnitudes, this means that we cannot,
without new assumptions, get a more accurate prediction than of this
order of magnitude of 2.5 GeV uncertainty in $ M_t$ and 5 GeV in $M_H$.
The uncertainty at present in the strong fine structure constant
$\alpha_3(M_Z)= 0.117 \pm 0.006 $ leads to an uncertainty in
our predictions of $\sim$ $\pm$ 2\%,
meaning $\pm$
3.5 GeV in the top quark mass. So our overall result for the top quark mass
is $ M_t = 173 \pm 5$ GeV.
Combining the
uncertainty from the Planck scale only being known in order of
magnitude and the $\alpha_3
$ uncertainty with the calculational uncertainty
in the vacuum stability curve, we get an overall uncertainty in
the Higgs boson mass of $\pm$ 9 GeV. So our Standard Model
criticality prediction for both the top quark and
Higgs boson pole masses is:
\begin{equation}
M_{t} = 173 \pm 4\ \mbox{GeV} \qquad M_{H} = 135 \pm 9\ \mbox{GeV}
\end{equation}

\section{The Ice-Water Analogue}

As we have just seen, the numerical coincidence is quite
impressive---we get very close to the experimental top quark mass!---but
the story about the time machine
paradox and the miracle may not a priori be too convincing.
It is therefore important to explain
that the multiple point criticality principle, which
is really equivalent to the requirement of degenerate vacua, is
rather closely analogous to a phenomenon happening in well-known and
well-established physics: We refer to the fine-tuning of
the temperature in a microcanonical ensemble, such as a mixture of
ice and water in a thermally isolated container.

Instead of arguing in terms of the time machine philosophy a very similar
sort of physics, and to tell the truth also one violating the principle
of locality, may be formulated in analogy to
such a microcanonical ensemble.
An action of the non-local type we have used above, and which represents
the very mild violation of locality (or causality) needed, is simply a highly
nonlinear function of reparameterisation invariant space-time integrals
$I_i \stackrel{def}{=} \int d^4x\sqrt{g(x)} {\cal L}_i(x)$
of field functions ${\cal L}_i(\phi, \partial \phi, ...)$
that could, by their symmetry properties, have
been Lagrangian densities \cite{book}.
Now we are interested in vacua and can thus consider
the Feynman path
integral with time taken purely imaginary;
then the configurations dominating the functional integral
should be the ones relevant for the vacua having the lowest
energy densites.
For our highly nonlinear (non-local) action,
the configurations dominating the functional integral
would be those with least Euclidean action;
in general these would correspond to
certain specific combinations of the
values for the various $I_i$.
At least it would be very usual to have some significant minima in the
non-local action as a function of the space-time integrals---the  $I_i$.
It would therefore often be a very good approximation to replace
the exponentiated action, $\exp(-S_{nl}(I_1,I_2,...,I_n))$,
by a product of delta functions
and, perhaps, still some extra factor multiplying them.

Remembering that the Euclidean path integral is formally
of the same type as a partition function for a statistical mechanical
model (e.~g.\ field theory), the appearance of delta functions brings
microcanonical ensembles to mind. Indeed we do believe that our
(mildly) non-local
actions are very likely
to be well approximated by
some microcanonical ensemble; or rather a generalization of
the microcanonical ensemble to an ensemble in which not only the
energy has a fixed value (the usual microcanonical ensemble), but also
other extensive quantities take on specified values (corresponding to
them being fixed by delta functions, under the integration leading to the
partition function).

Now there is a well-known mechanism whereby
microcanonical ensembles, or generalizations thereof,
lead to the appearance of more than one phase and the
fine-tuning of an intensive quantity: all very much analogous to the
consequences of the mild non-locality
for which we have already argued.
A very familiar example of a microcanonical ensemble is a certain number
of water molecules, in a multilayered plastic bag
that provides a thermally isolated system,
which has got a well defined and pre-specified amount of energy E.
Strictly speaking, for this system E should include the
energy spent in expanding the bag into the atmosphere; so
we should really use the enthalpy, $E_{\rm{water\ molecules}} + PV$,
where V is the (variable) volume and P is the fixed pressure.
Since it is not so practical to calculate directly with a microcanonical
ensemble, the standard procedure
is to approximate
the microcanonical ensemble with a canonical one. This procedure can
be formally performed by writing the delta function, specifying the
value of the Hamiltonian (really the enthalpy) H as equal to the
imposed energy E, as the Fourier transformation of a temperature
distribution. That is to say we write the delta function
\begin{equation}
\delta (H-E) = \int d\beta \exp(i\beta (H-E))
\end{equation}
and insert this Fourier expansion into the microcanonical partition
function
\begin{equation}
Z_{micro,E} = \int d{\bf q} d{\bf p} \delta (H({\bf q},{\bf p}) -E)
\end{equation}
Then  $Z_{micro,E}$ is expressed as an integral over the single
inverse temperature variable $\beta$ =$1/(kT)$ of a canonical partition
function  $Z(\beta)$
\begin{equation}
Z_{micro,E} = \int d\beta \; Z(\beta)\exp(-iE\beta)
\label{betaint}
\end{equation}
where
\begin{equation}
Z(\beta) = \int d{\bf q} d{\bf p} \exp(i\beta H({\bf q},{\bf P})).
\end{equation}
Now for large systems---say macroscopic systems---the integral
(\ref{betaint}) will be dominated by a very narrow range of $\beta$-values.
So, having
knowledge of the canonical ensemble, we may determine the
value of $\beta$ around which this very small significant $\beta$-interval
is centered, by finding that value which gives the prescribed value
$E$ for the average $<H>_{\beta}$ in the canonical ensemble.
Now, for our example of water molecules, it is well-known that
the energy or rather the enthalpy $<H>_{\beta}$ has a
$\theta$-function singularity, i.~e.~a
jump as a function of $\beta$ or, equivalently,
of the temperature $T$. So while it is often possible to find a
one-to-one correspondence between $E$ and
$\beta$---in fact $<H>$ is a monotonic function of $\beta$---there is
an interval of $E$-values which corresponds to the single
$\beta$-value at which the jump occurs.
In this interval the microcanonical ensemble must
in reality represent a system of two phases: in one region of the
container there is water, in another region there is ice.
The jump, of course, occurs at the
freezing temperature and is due to the finite
latent heat of fusion. Due to this latent
heat, the enthalpy of the water at the freezing point is higher
than that of the ice at the same temperature.

The absolutely crucial point for our purpose is that if
E is prescribed to be
in the gap at the jump discontinuity,
then the temperature, or equivalently $\beta$,
must be equal to the freezing temperature. That is to say a
very special temperature is obtained
{\em without fine-tuning} the energy (or enthalpy); but
rather only by prescribing it to take a value
within the jump discontinuity. The latter, however, does not
have to be fixed with any great precision,
but can be done just very crudely.
So one gets a fine-tuning---namely of the temperature or $\beta$---for free,
in the sense that nothing has to be fine-tuned at the outset!
The fact that in English there is a special word---slush---for the mixture
of ice and water which is really a microcanonical ensemble, and that
this slush is also used to fix the zero point on
the Celsius temperature scale,
indicates how easy it must be to fine-tune the
temperature by means of such a microcanonical ensemble.

\begin{figure}
\vspace{7.5cm}
\caption{Illustration of a graphical computation of the value of the
artificially introduced dummy (intensive) variable $\beta$.
Notice that a whole range of $<I_i>$-values gives
the same $\beta$-value as a solution, namely the value
$\beta_{crit}$ which
corresponds to the inverse temperature at the freezing
point in the ice-water analogy}
\label{fig:jump}
\end{figure}

We apply this phenomenon to the analogy in which the extensive
variables---the energy in the microcanonical ensemble---correspond
to quantities like the $I_i$ (integrals over the space-time manifold),
and the intensive quantities---the inverse
temperature in the canonical ensemble---are the effective coefficients
to the $I_i$ in an action linear in the $I_i$ and thereby a local
approximation to the true action; i.~e.~the intensive quantities are really
the coefficients in the action---the coupling constants. Then the remarkable
fine-tuning coming for free, in this case,
is that the coupling constants get fixed at those
values for which there is some first order phase transition. These are
precisely the values at which several vacua are degenerate; so this analogy
is very useful for us in that it predicts that vacua should be degenerate.

In our analogy the extensive variable $I_i$ corresponds to the
Hamiltonian H in a microcanonical ensemble. This is
illustrated in figure \ref{fig:jump}, where we can think
of $<I_i>$ as taking on a prescribed value $I_{i,\rm{fixed}}$
just as the average of the Hamiltonian $<H>$ must equal the
prescribed energy E. At the phase transition the intensive
variable (coupling constant) $\beta$ takes on the value
$\beta_{\rm{crit}}$, corresponding to a range of values
$\Delta I_i(\beta_{\rm{crit}})$ for the extensive variable.

This analogy throws some light on the reason for believing
that the values of the Higgs field in the two
degenerate vacua should differ by a
large amount. Suppose, for example,
that the integral of the Higgs field squared over all space-time is
the extensive quantity being fixed (like the energy in the microcanonical
ensemble):
\begin{equation}
I_1=\int d^4x \, |\phi(x)|^2
\label{I1}
\end{equation}
Then in order that it shall be likely that a randomly
prescribed value $I_{1,\rm{fixed}}$ for this integral
should just sit in the gap between the values in the two vacua
\begin{equation}
I_1(\phi_{\rm{min}\; 1}) \quad < \quad I_{1,\rm{fixed}}
\quad < \quad I_1(\phi_{\rm{min}\; 2})
\label{I1random}
\end{equation}
this gap must be large, meaning that the values of the
square of $\phi$ in the two vacua must differ appreciably. With
the Planck scale taken as the fundamental scale w.r.t.\ which this
``appreciableness '' is to be judged,
this condition becomes
\begin{equation}
\phi_{\rm{min}\; 2}-\phi_{\rm{min}\; 1}
\quad \simeq \quad M_{\rm{Planck}}
\label{appreciable}
\end{equation}
Otherwise the whole mechanism will be unlikely to work at all.
This is analogous to the fact that if the latent heat
of fusion for water is not large (in some
sense connected with the expected distribution of the prescribed
energy) then it will be very unlikely that slush should form;
the energy will usually either be too large so that only water
is formed, or too small so that only ice is formed. It is only
with a large latent heat and strong first order
phase transition that it becomes very likely to find slush.
In this analogy we do indeed see that the Planck scale difference
between the Higgs field values in the two vacua must be expected, for the
idea to be likely to work at all. Thus we see that
our model can really predict the top quark mass as
well as the Higgs boson mass.

\section{Fine Structure Constants, also from Multiple Point Requirement}
\label{sec:fine}

If we take the speculation seriously that, due to non-locality
or for other reasons, the vacuum should occur in more than one version
with the same energy density, then
it is not necessary  for there to be just
a couple of degenerate vacua: there could very likely be many.
The whole multiple point criticality
idea really developed out of earlier work \cite{picek}
aimed at predicting the three fine structure constants of
the Standard Model.
This work may, in the light of the above ideas,
be interpreted \cite{glasgowbrioni} as meaning that we look for degenerate
vacua in a model characterized by the following two extra assumptions:

1) There is a fundamental regularization, in the sense
that at small distances the laws of physics are truly different from
an ideal continuum quantum field theory. We effectively assume that
this regularization is provided by
a truly existing lattice; so that, for example, the
Yang Mills fields are fundamentally represented by link variables
taking values in the group, rather than by continuum fields.
Really, we hope that the precise way Nature is regularized does not
matter for the predictions of the critical couplings for which
the different vacua are degenerate. This hope has some support in the
work by Laperashvili \cite{laperashvili} and from some crude
estimates we have made.
So we expect the values of the fine structure constants,
for which phases with confinement already at the regularization scale
coexist with phases having confinement only at much
longer distance scales or not
at all, to be roughly independent of the precise method of regularization.
We also assume, in accordance with
our philosophy of taking Planck scale units as the fundamental
units, that the effective cut-off is provided by
the Planck energy: $\Lambda_{\rm{regularization}} = M_{\rm{Planck}}$.

2) At the Planck scale, or rather a
very moderate factor below it,
the Standard Model gauge group is extended, in very
much the same way as grand unified SU(5) is often assumed; it is just
that we assume another gauge group G, namely what we call anti-grand
unification (AGUT) in which G is $ SMG^3$ or $ SMG^3\times U(1)_f$.
Here $SMG= S(U(2)\times U(3))\approx  U(1)\times SU(2) \times SU(3) $
is the gauge group of the Standard Model
(SMG stands for \underline{S}tandard
\underline{M}odel \underline{G}roup).
This means that we assume that, close to the Planck scale, there are
36 or 37 types of gauge particle, instead of the 12 of the Standard
Model (dim(SMG) = 12), in such a way that there are 12 for each generation.
Roughly one can say that each generation gets its own photon, its
own $W^+$, its own $W^-$, etc. and then there is (perhaps) an extra
$U(1)$ gauge particle (called the $U(1)_f$ particle),
which is supposed not to matter so much for
deducing the values of the fine-structure constants
but is relevant for the fermion mass problem
discussed in the next section.

Of course we must also assume that, by means of Higgs fields or
some other mechanism,
the extended group $ SMG^3\times U(1)_f$ breaks down to the phenomenologically
observed Standard Model Group $SMG$, identified as the diagonal subgroup
of the $SMG^3$ = $ SMG \times SMG \times SMG$ group. The diagonal subgroup
is the group consisting of triplets $(u,u,u)$ of three identical elements
belonging to the abstract group $SMG$.
We then add the assumption that, for some reason, the
parameters in the laws of Nature get adjusted in such a way as to
make several, or rather as many as possible, vacua become
degenerate, in the sense of having the same energy density.

In practice lattice gauge theory calculations are usually
made in Euclidean space, i.e. with time taken as
purely imaginary. This formulation is well suited to study the
vacuum, which is by definition a ground state. If several
states of the vacuum with the same energy density occur, it will
show up in the Euclidean calculation as a transition
point---a phase transition point---as a function of some coupling
constant (a fine structure constant say); so that on one side
of the transition
the vacuum has one structure and on the other side another one.
The Euclidean calculation is dominated by the lowest energy density
state. Now if there are, say, two vacuum states with very
closely similar energy
densities, it will happen that for one value of some coupling
in the neighborhood (of that value for which the two vacua
are exactly degenerate) one or the other of the two vacua
will be the truly lowest energy one. Therefore the structure and
properties of the Euclidean calculation, say Monte Carlo computations,
will make a jump. There will be a phase transition in the
Euclidean partition function, at just the value of the
parameters for which two vacua are degenerate. If one wants to have
several degenerate vacua, it will in the Euclidean calculation correspond
to a multiple point at which several phases meet.

Now, in principle, we choose to calculate
how the parameters are adjusted
in the Euclidean lattice theory, so as to make
as many different phases as possible meet
for a single set of parameters. There will a priori be some
surface in the space of parameters along which this set of
phases meet. We take it, as our model, that Nature realizes
some point on this surface; then we suggest, supported
by some speculative calculations,
that the fine structure constants observed in the
continuum limit are
approximately the same all over this surface. Since we also suppose that
such continuum couplings at the phase transition points are
roughly independent of the type of regularization used, we
should be able to predict values for the fine structure
constants that are rather
stable with respect to the details of the calculation.
These numbers are to be identified
with the running fine structure constants, where the renormalisation point
is taken as the fundamental (Planck) scale.

In order to perform this program we have first to find out where we
can get as many phases as possible to meet. We should, in principle,
search in the space of all the possible parameters for the lattice theory
with our complicated group $SMG^3\times U(1)_f$; actually infinitely many
parameters because we have effectively assumed that the lattice really
exists.
For groups such as SU(2) or SU(3),
phase diagrams with two parameters already exist
in the literature \cite{lattice}, from which one immediately sees
that there is a point where three phases meet: one
described as being in the ``Coulomb phase'' at the lattice
scale and first confining at longer distances, one that
confines already at the lattice scale, and finally one phase
which, with respect to the continuum part of the group, is
in a similar ``Coulomb phase'' but confines at
the lattice scale with respect to the centre of the group, $Z_2$
or $Z_3$ respectively. There is a triple point where
the three phases meet for SU(2) or SU(3). We should
also remark that,
even away from the triple point,
the phases may merge into each other so that they are
not really separated; but that does not matter for our purposes,
since we are just interested in the existence of
a multiple point. By choosing
the couplings---lattice action
parameters---corresponding to this triple point for each of the three
$SU(3)$ cross-product factors in the AGUT group $SMG^3$
and the form of the action so as to have
no interaction between the three
$SU(3)$ factors, we obtain a set of parameters such that
we can achieve any combination of phases for the three $SU(3)$ factors
just by infinitesimal shifts of these parameter-values. Thus we can,
in this what we could call factorized way, obtain a multiple point
for the whole group $SU(3)^3$
where $3^3$ phases meet.
By analogy, it is also expected
that the different types of non-abelian group
in the AGUT group can be adjusted
independently to reach the multiple point for each type
separately; thereby  a multiple point for the whole cross product
is achieved, with a number of phases meeting equal to the
product of the number for each type alone. For the abelian groups,
however, we expect that the maximal number of phases meeting will
be achieved in a more complicated way. We do not have a completely
safe calculation of how many phases
can meet when
the abelian part of the group is taken into account, but we
estimate that, at least approximately, the meeting of the highest number
of phases is achieved in what we call a ``hexagonal'' scheme.
In this scheme, we have terms in the lattice action involving
more than one of the $U(1)$ groups in the cross-product at a time.
While there is no possibility for an interaction
of this type between the gauge
fields of non-abelian groups, $SU(2)$ and $SU(3)$,
there is such a possibility for the
abelian groups:
in a continuum formulation for $U(1)$ groups, there is an
allowed term in the Lagrangian density of the
form $F_{\mu\nu}^{Peter}(x) F^{Paul\; \mu\nu}(x)$.  Since even
$F^{Peter}_{\mu\nu}(x)$ is gauge invariant for the abelian case,
we are allowed to
construct this type of interaction between different
$U(1)$-gauge fields;
a corresponding
non-abelian term would not be gauge invariant.
Here the names $Peter$, $Paul$ etc. are
used to distinguish the different cross-product factors in G---in
this case the $U(1)$ factors.

The ``hexagonal'' form of the total lattice action for several
$U(1)$ groups contains various interaction terms,
which are lattice variants
of $F_{\mu\nu}^{Peter}(x)F^{Paul\; \mu\nu}(x)$ or more complicated
lattice terms involving several of the $U(1)$ groups, arranged so that
the total lattice action has a ``hexagonal''
symmetry.
This symmetry can, very abstractly, be
identified with the group of a hexagonally symmetric lattice in
a three dimensional space.
This symmetry property of the action at
the proposed multiple point reflects a group of rotations---or
at least linear  transformations ---in the covering group ${\bf R}^3$
of the group $U(1)^3$ contained in the group $SMG^3$. It is this
discrete group of rotations (when the action is used to define the
metric in the covering group space) which has the same structure as
the group of symmetries of a hexagonal lattice
in the covering group. By choosing a lattice action to have
symmetry like this hexagonal one,
the phases transformed
into each other under the symmetry operations are forced to
meet
at the proposed multiple point.
In this way one may relatively easily get a lot of phases to meet.
We have
not investigated carefully, yet, the effect on our results of the extra
abelian group $U(1)_f$, but we imagine that it essentially does not
mix with the other $U(1)$ factors and that we
can ignore it, as far as the diagonal
subgroup couplings are concerned.

At the multiple point where most phases meet, as estimated above,
we then evaluate the diagonal subgroup couplings. This is done
mainly by analytical estimates,
using computer generated data taken from
the literature on
the critical couplings for the groups
$SU(3)$, $SU(2)$, and $U(1)$ as input.
For the non-abelian groups
one gets, in first approximation, the following
simple expression for the
diagonal subgroup fine structure constant $ \alpha_{diag}$:
\begin{equation}
\frac{1}{\alpha_{diag}} = \frac{1}{\alpha_{Peter}} +
\frac{1}{\alpha_{Paul}} + \frac{1}{\alpha_{Maria}}
\end{equation}
where $\alpha_{Peter}$, $\alpha_{Paul}$ and $\alpha_{Maria}$
are the fine structure constants for the corresponding three
cross-product factors in $SMG^3$; at the proposed multiple
point their values are taken to be those at the
triple point for a single $SU(2)$ or
$SU(3)$ group. For the abelian
$U(1)$ groups the story is more
complicated and we rather get in first approximation:
\begin{equation}
\frac{1}{\alpha_{1\; diag}}= \frac{6}{\alpha_{crit}} \;\;\;\;
\mbox{(first approximation)}
\end{equation}
but with some corrections
(due to more complicated terms in the
action needed to ensure the multiple point)
the factor $6$ becomes
$6.8$.

Finally we obtain our predictions of the fine structure constants
at the Planck scale:
\begin{equation}
1/\alpha_1(M_{\rm{Planck}}) = 52\; (60)\; \pm 5\;\;\;\;
1/\alpha_2(M_{\rm{Planck}})=
48 \pm 6 \;\;\;\; 1/\alpha_3(M_{\rm{Planck}})   = 56 \pm 6
\end{equation}
where the number in brackets is the predicted
value of the $U(1)$ fine structure
constant after making the corrections beyond the first approximation
referred to above.
Extrapolation of these numbers to the the weak interaction scale,
using the following Standard Model
renormalisation group equations for the gauge coupling constants $g_i$,
\begin{equation}
16\pi^2\frac{dg_i}{d\ln\mu} = b_ig_i^3 ; \qquad \qquad \rm{where} \qquad
b_1 = \frac{41}{6}, \quad b_2 = -\frac{19}{6}, \quad b_3 =-7
\label{rgegauge}
\end{equation}
leads to
\begin{equation}
1/\alpha_1(M_Z)= 96 \;(103)\; \pm 5 \;\;\;\;
1/\alpha_2(M_Z) = 32 \pm 6 \;\;\;\;
1/\alpha_3(M_Z) = 16 \pm 6.
\end{equation}
There is agreement
with the ($\overline{MS}$) experimental numbers \cite{pdg95}
\begin{equation}
1/\alpha_{1, exp}(M_Z) =  98.29 \pm 0.09 \;\;\;\; 1/\alpha_{2,exp}(M_Z) =
29.61 \pm 0.06 \;\;\;\; 1/\alpha_{3,exp}(M_Z)= 8.55 \pm 0.44
\end{equation}
well within
the theoretical uncertainties,
which we have crudely estimated. They include uncertainties from
the Monte Carlo data used of the order of 5\% of the inverse fine structure
constants at the Planck scale. Our several corrections suffer from
uncertainties of the same order of magnitude. For the $U(1)$ coupling
there is still some uncertainty as to precisely which phases
meet at the proposed multiple point; in particular the question
as to whether some
of the discrete subgroups may or may not confine separately.

\section{Quark and Lepton Masses and Mixings compatible with
\protect\boldmath$SMG^3\times U(1)_f$ }
Given its success in predicting the
values of the Standard Model gauge coupling
constants, one might ask whether the $SMG^3$ model might also provide
some understanding of the values of the Standard
Model Yukawa coupling
constants and the quark-lepton mass problem.  Indeed the
broken chiral gauge quantum numbers of the quarks and leptons,
under the symmetry groups $SMG_a$ (a=1,2,3), distinguish between the
three generations and have the potential to generate the fermion mass
and mixing hierarchy \cite{book}. This hierarchy corresponds to
different degrees of suppression for various transitions from
right-handed to left-handed fermion states, each of which carry different
$SMG^3$ quantum numbers. The most promising way of explaining
these mass suppression factors is in terms of the partial
conservation of such chiral flavour quantum numbers \cite{corfu}.
In fact the mass gaps between the three fermion generations are
readily explained by the broken $SMG^3$ gauge quantum numbers.
Unfortunately, however, it is not possible to explain all the mass
splittings within each generation, such as the ratio of the top
and bottom quark masses, in this way. The $SMG^3$ model
inevitably predicts \cite{fln:np1}:
\begin{equation}
m_u m_c m_t \: \simeq \:m_e m_{\mu} m_{\tau} \: \leq \: m_d m_s m_b
\end{equation}
So, together with Gerry Lowe, we were led to extend the gauge group
by an extra abelian flavour group factor $U(1)_f$. If we maintain the
feature of the anti-grand unified model that the irreducible
representations of the Standard
Model are not combined into larger irreducible
representations of the extended gauge group (intuitively combining
irreducible representations together generally invites
unwanted mass degeneracies), then the $SMG^3 \times U(1)_f$
model is the only non-trivial anomaly-free extension of $SMG^3$
with no new fermions. Furhermore
the $U(1)_f$ charges are essentially unique:
\begin{equation}
\left( \begin{array}{ccccc}
d_L \;&\; u_R \;&\; d_R \;&\; e_L \;&\; e_R \\
s_L \;&\; c_R \;&\; s_R \;&\; \mu_L \;&\; \mu_R \\
b_L \;&\; t_R \;&\; b_R \;&\; \tau_L \;&\; \tau_R
       \end{array} \right)
=\left( \begin{array}{ccccc}
0 \;&\;  0 \;&\;  0 \;&\; 0 \;&\;  0 \\
0 \;&\;  1 \;&\; -1 \;&\; 0 \;&\; -1 \\
0 \;&\; -1 \;&\;  1 \;&\; 0 \;&\;  1
       \end{array} \right)
\label{eq:u1soln}
\end{equation}
We constucted a computer program to search over various
mass matrix scenarios, using partially conserved $SMG^3\times U(1)_f$
gauge quantum numbers, and found that the extended model could
naturally accomodate the fermion mass spectrum and mixing angles up to
unknown factors of order unity \cite{fln:np1}.

Very recently---and actually after the meeting  on Corfu---we
have developed, together with Douglas Smith, the type of fit
that the computer seemed to have reached into an
explicit representation, as a model
in terms of some supposed Higgs
fields breaking the group $SMG^3\times U(1)_f$ down to the Standard Model
Group $SMG$, and finally  to $SU(3)\times U(1)$
by the Weinberg-Salam Higgs field $\phi_{WS}$. The
Higgs field $\phi_{WS}$ has, in
our model, to take highly nontrivial quantum numbers under the group
$SMG^3\times U(1)_f$. Of course for the subgroup of it which is the usual
Standard Model Group, the quantum numbers of $\phi_{WS}$
must be the well-known weak hypercharge $Y/2=1/2$, doublet under SU(2)
and singlet under SU(3).
The Standard Model Higgs field, of course, satisfies the usual
charge quantisation
rule \cite{book}:
\begin{equation}
\label{SMchqu}
\frac{Y}{2}+\frac{1}{2}\rm{``duality"}+\frac{1}{3}\rm{``triality"}
	\equiv 0\pmod{1}
\end{equation}
The most natural
way to incorporate this charge quantisation rule in the
$SMG^3\times U(1)_f$ model is to assume that such a rule
holds for each component $SMG_a$. If we assume that we only
have singlet or fundamental  matter (fermion and scalar) field
representations for all the non-abelian
gauge groups $SU(3)_a$ and $SU(2)_a$,
the charge quantisation rules analogous to eq.(\ref{SMchqu})
determine the non-abelian representations from the $U(1)_a$
charges $y/2|_a$. So the three weak hypercharges  $y/2|_a$ and
the $U(1)_f$ flavour charge $Q_f$ together completely specify
a matter field representation.

We imagine that the group $SMG^3\times U(1)_f$ is spontaneously broken
down to the Standard Model Group $SMG$ by (essentially) four Higgs field
vacuum expectation values
(VEVs) called say: S, W, T and $\xi$ in units of the Planck mass.
These Higgs fields VEVs respect the usual SMG and hence have
weak hypercharges $y/2|_a$ satisfying
\begin{equation}
Y/2 = y/2|_1 + y/2|_2 + y/2|_3 = 0
\end{equation}
The biggest vacuum expectation value
is taken by the field S with quantum numbers
\begin{equation}
(y/2|_1,y/2|_2,y/2|_3,Q_f)|_S = (1/6,-1/6,0,-1),
\end{equation}
but that is supposed to be just of order unity in fundamental
scale units, so that matrix elements are in reality not at all
suppressed by this VEV, i.e. we take S = 1 and
act as if the group is totally broken down
as far as this field could achieve.
Since the Higgs field S gives
no suppression, it is hard to detect in a phenomenological fit
of the fermion mass spectrum. Consequently there is an
ambiguity in guessing the quantum numbers of the other VEVs, namely
W,T,and $\xi$. Instead of proposing for $\xi$ say
$ (y/2|_1,y/2|_2,y/2|_3,Q_f)|_{\xi}$ =$ (1/6,-1/6,0,0)$ we could as
well use, instead of $\xi$ itself, a combination of $\xi$ and S and
thus declare that really the quantum numbers of $\xi$ were
$ (y/2|_1,y/2|_2,y/2|_3,Q_f)|_{\xi}$ = $( 0,0,0,1)$. Modulo such extra
factors of $S$ we were led to the following proposal for the quantum
numbers of the Higgs field expectation values:
\begin{eqnarray}
(y/2|_1,y/2|_2,y/2|_3,Q_f)|_W = (0,-1/2,1/2,-4/3) \\
(y/2|_1,y/2|_2,y/2|_3,Q_f)|_T =  (0,-1/6,1/6,-2/3)  \\
(y/2|_1,y/2|_2,y/2|_3,Q_f)|_{\xi} = (1/6,-1/6,0,0)
\end{eqnarray}
The quantum numbers of the Weinberg-Salam Higgs field in our proposal are:
\begin{equation}
(y/2|_1,y/2|_2,y/2|_3,Q_f)|_{\phi_{WS}} = (0,2/3,-1/6,1)
\end{equation}

With these quantum numbers one can evaluate the combination  of Higgs fields
needed to provide nonzero values for the various matrix elements in the three
mass matrices, namely for the quarks with electric charge $Q=2/3$ called
$M_u$, the one for the quarks with $Q=-1/3$ called $M_d$ and finally for
the charged leptons $M_l$. Strictly speaking there are several possible ways
to produce the same quantum numbers, but usually one of those
ways is obviously dominating, since the other ones are too suppressed.
At least this is true if one uses our fitted values: (S = 1) , $T\simeq 1/12$,
$W\simeq 1/6$, and $\xi\simeq 1/10$. In this way the
following quark mass matrix structure was obtained:
\begin{equation}
M_u \simeq \left ( \begin{array}{ccc}S^{\dagger}W^{\dagger}T^2
(\xi^{\dagger})^2
& W^{\dagger}T^2\xi
& (W^{\dagger})^2T\xi\\ S^{\dagger}W^{\dagger}T^2(\xi^{\dagger})^3
& W^{\dagger}T^2 & (W^{\dagger})^2T\\
S^{\dagger}(\xi^{\dagger})^3 & 1
& W^{\dagger}T^{\dagger}  \end{array} \right )v
\qquad
M_d \simeq \left ( \begin{array}{ccc}SW(T^{\dagger})^2\xi^2
& W(T^{\dagger})^2\xi
& T^3\xi\\ SW(T^{\dagger})^2\xi & W(T^{\dagger})^2 & T^3\\
SW^2(T^{\dagger})^4\xi & W^2(T^{\dagger})^4
& WT  \end{array} \right )v
\label{qmm}
\end{equation}
While for the charged leptons:
\begin{equation}
M_l \simeq \left ( \begin{array}{ccc}SW(T^{\dagger})^2\xi^2
& W(T^{\dagger})^2(\xi^{\dagger})^3
& (S^{\dagger})^2WT^4\xi^{\dagger}\\
SW(T^{\dagger})^2\xi^5 & W(T^{\dagger})^2 & (S^{\dagger})^2WT^4\xi^2\\
S^3W(T^{\dagger})^5\xi^3 & (W^{\dagger})^2T^4
& WT  \end{array} \right )v
\label{lmm}
\end{equation}
Here $v = \langle\phi_{WS}\rangle/\sqrt{2} = 174$ GeV.
Unknown coefficients of $\cal O$(1) in the matrix elements have been
ignored. A three parameter order of magnitude fit
\cite{fns;96}, using eqs.~(\ref{qmm}) and (\ref{lmm}) with S = 1 fixed,
successfully reproduces all the observed fermion masses and mixing
angles within a factor of two---see table \ref{bestfit}.

\begin{table}[h]
\caption{Best fit to experimental data. All masses are running masses at 1 GeV
except the top quark mass which is the pole mass.}
\begin{displaymath}
\begin{array}{|c|c|c|c|c|c|c|c|c|c|}
\hline
 & m_u & m_d & m_e & m_c & m_s & m_{\mu} & M_t & m_b &
m_{\tau} \\ \hline
{\rm Fitted} & 3.8 {\rm \; MeV} & 7.4 {\rm \; MeV} &
1.0 {\rm \; MeV} & 0.83 {\rm \; GeV} & 415 {\rm \; MeV} &
103 {\rm \; MeV} & 187 {\rm \; GeV} & 7.6 {\rm \; GeV} &
1.32 {\rm \; GeV} \\ \hline
{\rm Experimental} & 4 {\rm \; MeV} & 9 {\rm \; MeV} &
0.5 {\rm \; MeV} & 1.4 {\rm \; GeV} & 200 {\rm \; MeV} &
105 {\rm \; MeV} & 180 {\rm \; GeV} & 6.3 {\rm \; GeV} &
1.78 {\rm \; GeV} \\ \hline
\end{array}
\end{displaymath}
\begin{displaymath}
\begin{array}{|c|c|c|c|}
\hline
 & V_{us} & V_{cb} & V_{ub} \\ \hline
{\rm Fitted}  & 0.18 & 0.029 & 0.0030 \\ \hline
{\rm Experimental}  & 0.22 & 0.041 & 0.002 - 0.005 \\ \hline
\end{array}
\end{displaymath}
\label{bestfit}
\end{table}

\section{Conclusion and Consolation}

Based on two or three ``basic'' assumptions, we have been able to
predict an impressive number of parameters in the Standard Model.
The assumption which we studied most in the present article
is what we have often called the ``multiple point criticality
principle''; this means that the Universe shall allow for more than
one---really as many as possible---vacuum states with
(at least approximately) the same energy density. This assumption
could be understood in a model in which there is a very mild form
of non-locality, namely where the coupling constants are influenced
by averages over all space-time (the future as well as the past), and then
the degeneracy of the vacua appeared as a miracle needed to
make the time machine (of the mild form) be consistent.
Our cleanest prediction is that, given the top quark mass from
experiment to be $M_t = 180 \pm 12$ GeV, the Higgs boson
mass should be  $M_H = 149 \pm 26$ GeV. But with the
additional assumption that
the phase transition should be strongly first order,
so that this time machine mechanism should have
a good chance of being relevant,
we require that the difference
in the Higgs field vacuum expectation values
between the two phases should be rather big,
on the scale of the supposed fundamental units---Planck units.
This leads to the prediction of the top quark mass
as  $M_t = 173 \pm 5$ GeV and a more precise prediction of
the Higgs particle mass as
$M_H = 135 \pm 9$ GeV.

To get further predictions it was necessary to include
our second basic assumption, namely that there is
what is sometimes called anti-grand unification,
meaning we introduce the non-simple ``unifying'' group $SMG^3$
or its extension to
$SMG^3\times U(1)_f$.
Really this means that we assign a separate set of
gauge particles---with a mutual interaction structure just identical
to that of the set of 12
gauge particles in the Standard Model---for each of the three generations.
The $U(1)_f$ group contributes an extra
abelian
gauge particle, coupling to the right handed particles
in the second and third generations in a way allowed by the
anomaly cancellation
constraints. The latter requirement to a very large extent fixes
the $U(1)_f$ charges of all the fermions.

The main motivation for considering this rather specific
AGUT gauge group, broken down to the diagonal subgroup
of $SMG^3$ at the Planck scale is, of course, its
successful phenomenological predictions. However it
originated from some speculative ideas \cite{book}, involving the
``confusion'' mechanism in random dynamics and the
observation that the Standard Model Group SMG is very
skew---in the sense of having exceptionally few
automorphisms. In fact, perhaps most importantly, we can
characterise the group $SMG^3\times U(1)_f$
as the largest possible gauge group beyond the
Standard Model, satisfying the following principles:
\begin{enumerate}
\item
Each gauge field transforms the
already known 45 species of Weyl fields (3 generations with 15 in each)
nontrivially; thus we only look for a gauge group $G\subset U(45)$,
where $U(45)$ here stands for the group of
unitary transformations of these known 45 Weyl fields.
\item
There should be neither gauge anomalies nor mixed anomalies.
\item
The various irreducible representations of Weyl-fields for the
Standard Model Group are required {\em not} to be united under the full
group $G$, but rather to remain as separate
irreducible representations even under $G$.
\end{enumerate}

The predictions of the three fine structure constants
in the Standard Model, using
the principle of having degenerate
vacua and
our AGUT gauge group,
agree well with experiment---in fact
better than our calculational accuracy until now.
Finally we put forward a
model of the fermion masses, using our favourite gauge group
$SMG^3\times U(1)_f$,
and obtained
a rather good fit to the {\em order of magnitudes} of
the quark and lepton masses in terms of only three parameters.

All our results are compatible with a single model, in which there is
no significant new physics until
an order of magnitude below the Planck scale,
where one finds the scalar fields connected with our mass matrix fit
and the spontaneous breakdown of
the AGUT group to the Standard Model Group.
Even higher in energy there is some truly existing
regularization---but it is
speculated not to be so
significant which one, even the superstring could
be classified as a  useful regularization for our
purpose---and there will then be very likely lots of new physics.
Especially wormholes or baby universes are
expected to be present in the
space-time foam and cause the mild breaking of locality, which we
use in our model to argue for the degenerate vacuum hypothesis.
Again the details, of whether it is really baby universes or
wormholes that do this job or that locality is just
not a fundamental
principle of Nature, are unimportant for our predictions.

We note that both a vanishing cosmological constant,
$\Lambda_{eff}=0$, and strong CP conservation,
$\Theta_{QCD}=0$, are characterised as meeting points of phases
\cite{hamber,schierholz}. Thus mild non-locality, and its
affinity for co-existing phases, could also help to
explain the cosmological constant problem and the
problem of why $\Theta_{QCD}$
(and the weak vacuum angle $\Theta_{weak}$ which
is presumably not practically measurable) should be
so small. Then we can claim to have fitted
at least order of magnitudewise
{\em all} the parameters of the Standard Model, even including gravity,
in our approach. However we must admit to have taken the
Standard Model Higgs vacuum expectation value,
$<\phi> = 246$ GeV, from experiment. Our hope of
explaining the gauge hierarchy problem, in terms of
a vanishing Higgs mass at a phase boundary
\cite{glasgowbrioni,fln:np1}, fails for the
strongly first order phase transition we
have argued for in this paper.

If one accepts the success of our fit of the Standard Model
parameters as confirming our model, it is in a way sad:
its main consequence is that experiments will not find
any supersymmetric partners and, on the whole, very little new
physics at accessible energy scales. We simply predict a
Higgs particle of mass $M_H = 135 \pm 9$ GeV and then
essentially a desert up to the Planck scale. The presence of
other particles near the electroweak scale would contribute to
the renormalisation group beta functions and, in general,
would spoil our successful predictions of the top quark mass and
the fine structure constants.

Well: there are one or two suggestions for possible new
physics in our model, which could just have a
chance of being observed:

\begin{itemize}
\item
The Weinberg-Salam Higgs field in our AGUT scheme actually
belongs to a multiplet together with
coloured components. Such coloured scalar particles could
for symmetry reasons be, like the Weinberg-Salam Higgs field itself,
very light from a Planck scale point of view. Although this does not mean
that they must be light enough to be experimentally accessible in
the near future, there is at least a chance that they are.

\item
In our estimates of the fine structure constants, it seems
that discrete subgroups
of the gauge group play a non-negligible r\^{o}le and actually improve
the predictions somewhat. Now contrary to the continuum part
of the group---the Lie group part so to speak---a discrete group
gauge theory does not give rise to any massless particles in general.
With our principle of degenerate vacua, however, there is a special
chance that some ``low energy'' physics (compared to the Planck scale)
could result:
According to our multiple point criticality principle,
discrete subgroups are expected to lie
on the phase boundary between confining and Coulomb-like
degenerate vacua.
If for some reason the phase transition between
the vacua happened to be second order rather than first
order, then the flux strings, which really constitute the
physics of discrete gauge theories, would have just zero string tension.
The point is that in the confinement phase the
electric flux appears as strings, whereas in the
Coulomb-like phase the magnetic flux appears  as strings.
Since both should
continuously  reappear or disappear at a
second order phase transition, the tensions
must go to zero there.
If this indeed happened,
some flux strings from discrete subgroups
might be able to contribute forces between particles
charged with the discrete charges. This would be with
zero tension
from the Planck scale point view,
again giving a chance of being found
experimentally some day.
Very likely we would first see such effects as radiative corrections.
For instance a vertex could be corrected by two of the involved
particles having a flux string connecting them, while they separate
after having been locally created from the third particle.
Could this possibly be an idea for explaining the deviations
from the Standard Model predictions for the $Z^0 \rightarrow b +\bar{b} $
or $Z^0 \rightarrow c +\bar{c}$ decay modes
discussed at this meeting? The discrete subgroup
involved could easily couple differently to the different quark flavours,
because it does not necessarily have to be a subgroup of the
Standard Model Group itself but could be e.~g.~a subgroup of our AGUT
group.

\item
Really our caculations would not be spoilt, if the new physics at low
energy did not appreciably disturb the renormalisation group running
of the Higgs self coupling, the top Yukawa coupling
and the gauge coupling constants.
Extra scalar fields would contribute rather weakly
to the running of the gauge couplings.
Also, given the precedence of the Weinberg-Salam Higgs particle
coupling very weakly to many particles,
it is not out of the question that some scalar(s) could
couple so weakly to the matter fields as to
leave our predictions essentially undisturbed.

\item
Systems of particles decoupled from the ones we know
are of course not excluded, but they would
be harder
to see the closer they are to being completely decoupled.
\end{itemize}

 But, in general, we must admit that the confirmation of our model
would mean that there will be no
 new physics at accelerators on any reasonable time scale!

 In this ``sad'' scenario, one might foresee a future in which
 research on the fundamental scale forces
 could only be done by performing
 high accuracy investigations of the
 Standard Model parameters and deducing
 information about details of Planck scale physics.
If there was really a good theory, one might learn something
from any given amount of information rather independently
of whether it was taken at very high energy
or with high accuracy at more accessible
energies, provided it is not just a general consequence
of the low energy tail of the theory.
Indeed if our influence from the
future picture were true, then it might, to a very
weak extent, be possible
to see the future through the measurement
of the parameters of the Standard Model ! In a way we are already
on the way to read the future in the Standard Model parameters,
when we propose that
our model can be checked by investigating whether the
mass of the Weinberg-Salam Higgs
particle has the
borderline value $135 \pm 9$ GeV. If indeed it has, we are
seeing a bad prophecy for the future in the parameter $M_H$:
we shall all kill ourselves by a vacuum bomb!

But one could imagine that with a deep understanding of
the numbers one could---and this is perhaps rather
realistic---estimate
the degree of difficulty in making such a vacuum bomb, by
accurately determining the Higgs mass and thereby the gain in
energy per unit volume achievable by a transition to the new
vacuum state. The less the gain in energy,
the bigger a bubble of new vacuum has to be
before it can expand by itself. One would expect that
the easier it is to produce the critical bubble size, and thereby the
bomb, the easier it will be to get funds for
building an accelerator that could achieve it
and the sooner
will be the end. So a not completely unrealistic attempt to
predict something about the future from the Standard Model parameters
is actually at hand!

\section*{Acknowledgements}
We have pleasure in thanking George Zoupanos and his colleagues
for all their efforts in organising this very successful
School and for their great hospitality. Financial
support from the British Council and EC grant CHRX-CT94-0621
are gratefully acknowleded by CDF and HBN respectively.

\end{document}